\documentclass{article}
\usepackage[a4paper,margin=2cm]{geometry}
\usepackage{graphicx}
\usepackage[T1]{fontenc}
\usepackage{txfonts}
\usepackage{mathptmx}
\usepackage[pdflang={en-US},pdftex]{hyperref}
\usepackage{color}
\usepackage{booktabs}
\usepackage{textcomp}
\usepackage{footmisc}
\usepackage{multirow}
\usepackage{bbding}
\usepackage{subfig}
\usepackage{wasysym}
\usepackage{xcolor}
\usepackage{fontawesome}
\usepackage{microtype}
\usepackage{ccicons}
\usepackage{seqsplit}

\definecolor{darkgreen}{rgb}{0.0, 0.5, 0.0}
\definecolor{orange}{rgb}{0.84, 0.96, 0.52}
\definecolor{amber}{rgb}{1.0, 0.49, 0.0}
\definecolor{darkblue}{rgb}{0.03, 0.22, 0.43}
\definecolor{violet}{rgb}{0.29, 0.0, 0.51}

\newcommand{\gc}{\fontsize{11pt}{5pt}\selectfont{\textbf{\color{darkgreen}\raisebox{-2pt}{\faCheck}}}}

\newcommand{\rc}{{\fontsize{11pt}{5pt}\selectfont{\textbf{\color{red}\raisebox{-2pt}{\faClose}}}}}

\newcommand{\ec}{{\fontsize{11pt}{5pt}\selectfont{\textbf{\color{amber}\raisebox{-2pt}{\faExclamation}}}}}

\newcommand{\ap}{{\fontsize{11pt}{5pt}\selectfont{\textbf{\color{black}\raisebox{-2pt}{\faApple}}}}}

\newcommand{\an}{{\fontsize{11pt}{5pt}\selectfont{\textbf{\color{black}\raisebox{-2pt}{\faAndroid}}}}}

\newcommand{\sm}{{\textbf{\color{violet}\raisebox{-2pt}{\faCircle}}}}

\newcommand{\direct}{{\large \textbf{\color{darkblue} \Circle}}}
\newcommand{\semidirect}{{\large \textbf{\color{darkblue} \LEFTcircle}}}
\newcommand{\indirect}{{\large \textbf{\color{darkblue} \CIRCLE}}}

\def\plaintitle{Developing Accessible Mobile Applications with Cross-Platform Development Frameworks}

\def\emptyauthor{}
\def\plainkeywords{Screen reader, cross platform development, accessibility, mobile applications.}

\newcommand{\code}[1]{{\ttfamily\seqsplit{#1}}}

\makeatletter
\def\url@leostyle{%
  \@ifundefined{selectfont}{
    \def\UrlFont{\sf}
  }{
    \def\UrlFont{\small\bf\ttfamily}
  }}
\makeatother
\definecolor{linkColor}{RGB}{6,125,233}
\hypersetup{%
  pdftitle={\plaintitle},
  pdfauthor={\emptyauthor},
  pdfkeywords={\plainkeywords},
  pdfdisplaydoctitle=true, 
  bookmarksnumbered,
  pdfstartview={FitH},
  colorlinks,
  citecolor=black,
  filecolor=black,
  linkcolor=black,
  urlcolor=linkColor,
  breaklinks=true,
  hypertexnames=false
}

\begin{document}

\title{\plaintitle}

\author{Sergio Mascetti \and Mattia Ducci \and Niccol\'o Cant\`u \and Paolo Pecis \and Dragan Ahmetovic}
\date{Universit\`a degli Studi di Milano}

\maketitle

\begin{abstract}
This contribution investigates how cross-platform development frameworks (CPDF) support the creation of mobile applications that are accessible to people with visual impairments through screen readers.
We first systematically analyze screen-reader APIs available in native \textit{iOS} and \textit{Android}, and we examine whether and at what level the same functionalities are available in two popular CPDF: \textit{Xamarin} and \textit{React Native}.
This analysis unveils that there are many functionalities shared between native \textit{iOS} and \textit{Android} APIs, but most of them are not available in \textit{React Native} or \textit{Xamarin}.
In particular, not even all \emph{basic} APIs are exposed by the examined CPDF.

Accessing the unavailable APIs is still possible, but it requires an additional effort by the developers who need to know native APIs and to write platform specific code, hence partially negating the advantages of CPDF.
To address this problem, we consider a representative set of native APIs that cannot be directly accessed from \textit{React Native} and \textit{Xamarin} and show sample implementations for accessing them.
\end{abstract}

\section{Introduction}
When mobile developers write native code, they interact with the operating system (OS) through application program interfaces (APIs) exposed by the OS itself or by additional platform-specific development frameworks, which we label together as \textit{system APIs}.
Normally, each platform has its own set of system APIs, accessed through platform-specific programming languages.
Therefore, applications intended for multiple platforms generally need to be developed separately for each different platform.

To address this problem, cross-platform development frameworks (CPDF) wrap the system APIs for different platforms into their own APIs, which are consistent among different platforms and can be accessed through a single programming language.
Thus, using CPDF, developers can easily implement cross platform applications, writing the code only once.
This is particularly useful for mobile development since two distinct popular platforms exist, namely \textit{iOS} and \textit{Android}.
For example, both \textit{iOS} and \textit{Android} APIs define their own button view\footnote{A \emph{view} is the basic building block for creating user interface components such as buttons.}.
CPDFs define their own button implementations 
that are then converted to the actual native \textit{iOS} or \textit{Android} buttons.
Since CPDF wrap many of the commonly used system APIs, they have been shown to reduce projects' development and maintenance costs~\cite{andrade2015comparative}.


One problem with CPDF, however, is that some of the native APIs are not accessible through them.
The developers can still access these APIs, but they need to write \textit{platform-specific code}, which means that the same function needs to be written once for each target platform.
This is not only time-consuming, but it also requires the developer to know the native APIs, hence limiting the advantages of CPDF themselves.

Previous contributions uncover that many accessibility-related APIs are unavailable in CPDF, in particular those needed to interact with the screen reader~\cite{zhuang2013tradeoffs,cantu2018mathmelodies,mascetti2019wordmelodies}.
However, to the best of our knowledge, the problem of how CPDFs support the development of accessible applications has never been systematically analyzed.


To address this problem,
this paper presents three contributions: first, it reports a detailed comparison of mobile screen-readers APIs obtained through a systematical analysis of \textit{iOS} and \textit{Android} platforms documentation. This comparison shows that the two considered native platforms expose many functionally equivalent screen-reader related APIs.
The second contribution is the analysis of how CPDF wrap these native APIs. We show that \textit{React Native} and \textit{Xamarin}, two of the most adopted CPDFs, only wrap about half and one third of the screen reader-related APIs shared by \textit{iOS} and \textit{Android}, respectively. 
Even worse, the two considered CPDFs do not even expose all the \textit{basic} accessibility APIs (a definition of \emph{basic} accessibility API is provided in the following). 
%
The third contribution consists in showing how to access the APIs that are not wrapped by CPDF. Considering a representative set of APIs, we show examples for accessing them in \textit{Xamarin} and \textit{React Native}, with platform-specific code. 
A set of example apps, each available in \textit{iOS}, \textit{Android}, \textit{React Native} and \textit{Xamarin} is introduced in this paper and is available online\footnote{\url{https://ewserver.di.unimi.it/gitlab/public_accessibility_software/mobilescreenreadersapi}\label{fn:link}}.



\section{Background}
\subsection{Mobile Assistive Technologies and screen readers}
Mobile devices are currently accessible to many people with disabilities, including people with upper limb mobility impairments, people with low vision, and blind people.
To achieve this, mobile operating systems allow people with disabilities to personalize how they interact with the device, for example providing a \textit{magnifier} to people with low vision.
%
In this paper we focus on mobile assistive technologies for people who are blind or have a severe visual impairment and cannot rely solely on their residual sight to interact with the device.
In such cases the assistive technology commonly used is the screen reader, which verbally describes elements on the screen~\cite{kane2008slide}.
Both Android and iOS mobile platforms present screen reading software.
On \textit{iOS}, \textit{Voice Over} screen reader is part of the OS itself and third party solutions cannot be used. \textit{Android}, instead, exposes APIs that enable third parties to develop screen readers.
In practice, however, \emph{Talk Back} screen reader, provided by Google, is most commonly used\footnote{Note that \emph{Talk Back} is often distributed as a pre-installed application on \textit{Android}, but it is a standalone app, not part of the OS.}.
Hence in this paper we only take into account \textit{Talk Back} and \textit{Voice Over}.

Screen readers can render any application accessible, also those developed by a third party. In some cases an application can be (at least partially) accessible through a screen reader even though the application developer does not take explicit actions to enable screen reader accessibility.
However, creating fully accessible apps often requires intervention from the developers. For example the developer has to specify the alternative text for images so that the screen reader can read them aloud.
This form of intervention can be achieved though \emph{screen reader APIs} that developers can use to interact with the screen-reader or to personalize its behaviour.
The screen reader APIs of the two considered screen readers expose a number of shared functions, which characterize the fundamental non-visual interaction, but
they also differ for a number of system-specific features, for example \textit{VoiceOver} enables the use of a \textit{rotor} gesture which is unavailable on \textit{TalkBack}.

\subsection{Cross-Platform Development Frameworks}
Cross-platform development can be achieved with four approaches: web apps, hybrid apps, interpreted apps, and generated apps~\cite{xanthopoulos2013comparative}.
The first two approaches use web technologies and therefore may suffer from an overhead during interaction.
Instead, in the latter two approaches, native code is automatically generated to create the user interface, which yields better performance~\cite{willocx2015quantitative}, hence resulting in a better user experience.
Indeed these are the ones adopted by the most recent technologies, including \textit{Xamarin} and \textit{React Native}.

To the best of our knowledge, no peer-reviewed publication compares the diffusion of CPDF. However, according to a 2019 survey conducted by StackOverflow\footnote{\url{https://insights.stackoverflow.com/survey/2019\#technology-\_-other-frameworks-libraries-and-tools}}, \textit{React Native} emerges as the most popular CPDF, followed by \textit{Xamarin}. For this reason this paper focuses on these two technologies that we briefly introduce in the following.

\subsection{React Native}
\textit{React Native}\footnote{\url{https://reactnative.dev/}} is a CPDF based on the \emph{interpreted app} approach~\cite{xanthopoulos2013comparative}.
The programming language is JSX, which is an extension of JavaScript that allows the developer to specify user interface with a markup code similar to XML.
At compile-time this code is converted into bare JavaScript that is then interpreted at run time.
The user interface is also generated at run time by first creating an abstract representation of the views and then transmitting it to the \textit{React Native} library that eventually converts it into the actual user interface composed of native view elements.
Therefore, APIs related to user interface are exposed through the defined JSX syntax, while other system APIs are exposed in the form of JavaScript classes.

In case a system API is not exposed by \textit{React Native}, the developer can use two approaches, one based on \textit{native modules}, the other on \textit{native components}.
With \textit{native modules} the developer can define functions in platform-specific code and then call these functions from the JSX code.
\textit{Native components}, instead, allow the developer to specify views in platform-specific code and use them from the JSX code.
In both cases the developer needs to know native APIs and to write platform-specific code in native languages, \textit{i.e.}, \textit{Objective-C} for \textit{iOS} and in \textit{Java} for \textit{Android}. 

\subsection{Xamarin}
\begin{figure}[t]
	\centering\hspace{1em}
\includegraphics[width=0.6\textwidth]{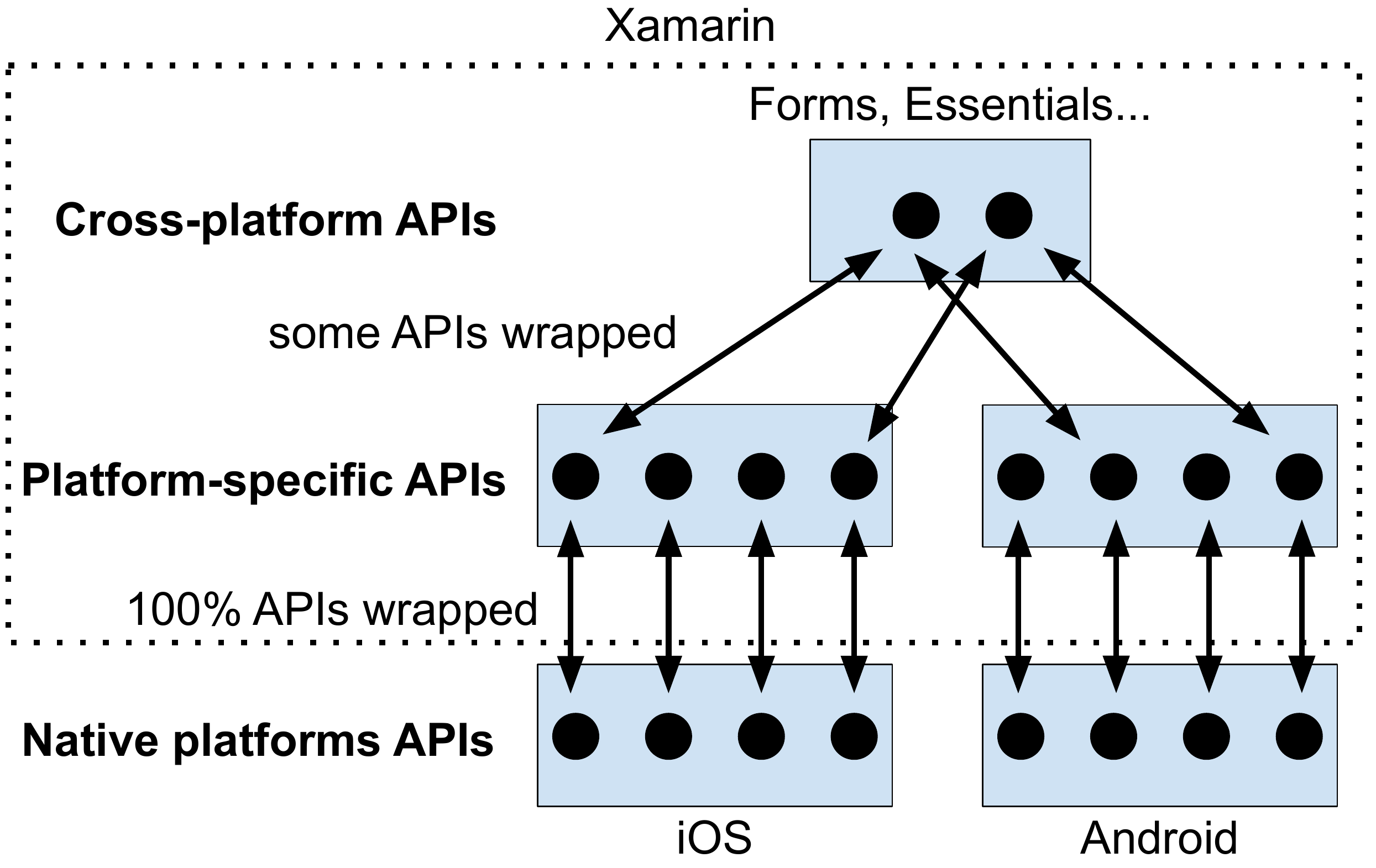}
	\caption{\textit{Xamarin} API mapping} 
	\label{fig:xamarin}
\end{figure}

\textit{Xamarin}\footnote{\url{https://dotnet.microsoft.com/apps/xamarin}} is a CPDF based on the \emph{generated app} approach~\cite{xanthopoulos2013comparative}. Applications are written in $C\#$ and then converted into native code. As depicted in Figure~\ref{fig:xamarin}, the framework wraps all native APIs from \textit{iOS} and \textit{Android} into \textit{platform-specific} APIs that are exposed to the $C\#$ code but are distinct for the two native platforms.
%
To allow the developer to write the code only once, \textit{Xamarin} also includes some libraries (\textit{e.g.}, \textit{Essentials} and \textit{Forms}) that remap platform-specific APIs into \textit{cross-platform} APIs.
For example, using \textit{Forms} the developer can insert a button view only once and this will result in inserting a native button in the target native platform (\textit{iOS} or \textit{Android}).
At run-time the right platform-specific APIs are used, depending on the actual platform where the code is running.
Without this remapping, adding a button view would require the developer to write the code twice, once for \textit{iOS} and once for \textit{Android}.

One problem is that these cross-platforms libraries only remap some of the platform-specific APIs.
To provide access to platform-specific APIs that are not remapped into cross-platform APIs, \textit{Xamarin} offers two approaches. One, called \emph{dependency service}, allows the developer to define a method that can be called from cross-platform code and whose implementation is specified in platform-specific code.
The second approach is based on \emph{custom renderer} and allows the developer to personalize cross-platform views defined in \textit{Forms}, including their aspect and behaviour.

One advantage with respect to \textit{React Native} is that accessing the platform-specific APIs the developers write code in $C\#$ only and do not need to use other programming languages; however, similarly to \textit{React Native}, the developers still need to know how platform-specific APIs work.

Note that in this paper we say that the platform-specific APIs that are not remapped into cross-platform APIs are not \textit{wrapped} (or \textit{exposed}) by \textit{Xamarin}; strictly speaking this is improper, because they are actually wrapped from native code to platform-specific libraries. However, since we are interested in cross-platform APIs, in analogy with the terminology that we use for \textit{React Native} we say that they are not \textit{wrapped} (or \textit{exposed}).



\section{Native Screen-reader API}
We conducted an analysis of screen reader APIs by considering the development documentation for \textit{iOS} and \textit{Android} platforms and by experimenting with the actual implementation, as described in the following of this paper.  
We created a taxonomy of the identified APIs based on the conceptual functionality they expose to the developer.
The result of this analysis is shown in Table~\ref{tab:concepts}, which lists the $25$ identified API functionalities and their availability in the four considered platforms (native \textit{iOS}, native \textit{Android}, \textit{React Native}, \textit{Xamarin}).
Tables~\ref{tab:nativescreenapi} and \ref{tab:crossplatformscreenapi} (see ``Supplementary Materials'' appendix) 
report additional information on how each API is implemented.

\begin{table*}[h!]
    \fontsize{7.5}{11}\selectfont
    \centering
    \renewcommand{\arraystretch}{1.2}
    \caption{Screen reader API functionalities and their availability in development platforms. Notation: {\gc}~available; {\rc}~not available; {\ec}~available with limitations; basic according according to \textit{Android} ({\an}) or iOS ({\ap}) documentation.}
    \label{tab:concepts}
    \begin{tabular}{|c|c|p{.65cm}|p{10.7cm}|c|c|c|c|}
    \hline
    \textbf{\rotatebox[origin=c]{90}{Category}} &
    \textbf{ID} &
    \hspace{.5em}\textbf{\rotatebox[origin=c]{90}{Basic in}} &
    \textbf{API functionality} &                              \textbf{\rotatebox[origin=c]{90}{iOS}} &
    \textbf{\rotatebox[origin=c]{90}{Android}} &
    \textbf{\rotatebox[origin=c]{90}{React N.}} &
    \textbf{\rotatebox[origin=c]{90}{ Xamarin }} \\ 
    \hline
    \hline
    \multirow{5}{*}[2.0ex]{\centering\rotatebox[origin=c]{90}{\hspace{-1cm}Acc. focus}} &
    $\#1$ & \centering\an &
    Specify which views should receive the accessibility focus &
    \gc & \gc & \gc & \gc \\[2pt]
    \cline{2-8}
    & $\#2$ & \centering\ap
    & Specify the accessibility focus order &
    \gc & \gc & \rc & \gc \\[2pt]
    \cline{2-8}
    & $\#3$ &
    & Assign the accessibility focus to a view &
    \gc & \gc & \gc & \rc \\[2pt]
    \cline{2-8}
    & $\#4$ &
    & Specify actions associated to accessibility focus-related events (e.g., a view acquires or loses the focus) &
    \gc & \gc & \rc & \rc \\[2pt]
    \cline{2-8}
    & $\#5$ &
    & Determine whether a view has the accessibility focus or which view has the accessibility focus &
    \gc & \gc & \rc & \rc \\[2pt]
    \hline
    \hline
    \multirow{5}{*}{\rotatebox[origin=c]{90}{\hspace{-.5cm}Text to ann.}} &
    $\#6$ & \centering\an~\centering\ap &
    Specify attributes that contribute to form the text-to-announce &
    \gc & \gc & \centering\ec & {\centering\ec} \\[2pt]
    \cline{2-8}
    & $\#7$ &
    & Programmatically define the text-to-announce &
    \rc & \gc & \rc & \rc \\[2pt]
    \cline{2-8}
    & $\#8$ & \centering\an &
    Use one view to describe another one &
    \rc & \gc & \rc{} & \gc{} \\[2pt]
    \cline{2-8}
    & $\#9$ &
    & Specify that a view should be announced when changed, even without user interaction &
    \rc & \gc & \gc{} & \rc \\[2pt]
    \cline{2-8}
    & $\#10$ &
    & Specify in which language the text-to-announce should be read &
    \gc & \gc & \rc & \rc \\[2pt]
    \hline
    \hline

    \multirow{7}{*}{\rotatebox[origin=c]{90}{\hspace{-.5cm}Explicit TTS}} &
    $\#11$ & \centering\ap &
    Read a text with the screen-reader TTS &
    \gc & \gc & \gc & \rc \\[2pt]
    \cline{2-8}
    & $\#12$ &
    & Be informed when the screen-reader finishes reading an explicitly provided text &
    \gc & \rc & \gc{} & \rc \\[2pt]
    \cline{2-8}
    & $\#13$ &
    & Customize screen-reader TTS speech features, like pitch, speed, etc... &
    \gc & \rc & \rc & \rc \\[2pt]
    \cline{2-8}
    & $\#14$ &
    & Read a text with non-screen-reader TTS (also works when screen-reader is not active) &
    \gc & \gc & \gc & \gc \\[2pt]
    \cline{2-8}
    & $\#15$ &
    & Detect whether the non-screen-reader TTS is reading &
    \gc & \gc & \gc & \rc \\[2pt]
    \cline{2-8}
    & $\#16$ &
    & Pause the non-screen-reader TTS &
    \gc & \rc & \rc & \rc \\[2pt]
    \cline{2-8}
    & $\#17$ &
    & Customize non-screen-reader TTS speech features, like pitch, speed, etc... &
    \gc & \gc & \gc & \gc \\[2pt]
    \hline
    \hline
    \multirow{4}{*}{\rotatebox[origin=c]{90}{\hspace{.15cm}Acc. tree}} &
    $\#18$ & \centering\an &
    Aggregate multiple  views into a single accessible element &
    \rc & \gc & \gc{} & \gc{} \\[2pt]
    \cline{2-8}
    & $\#19$ &
    & Decompose a view into multiple accessibility elements &
    \gc & \gc & \rc & \rc \\[2pt]
    \cline{2-8}
    & $\#20$ &
    & Get the parent accessible element &
    \gc & \gc & \rc & \rc \\[2pt]
    \hline
    \hline
    \multirow{4}{*}{\rotatebox[origin=c]{90}{\hspace{-.8cm}Miscellaneous}} &
    $\#21$ & &
    Detect whether screen-reader is active &
    \gc & \gc & \gc & \rc \\[2pt]
    \cline{2-8}
    & $\#22$ & \centering\an
    & Support navigation by specifying which views are headers or panes &
    \rc & \gc & \gc & \rc \\[2pt]
    \cline{2-8}
    & $\#23$ & \centering\ap
    & Define how to respond to user actions that are only available when the screen reader is active &
    \gc & \rc & \gc{} & \rc \\[2pt]
    \cline{2-8}
    & $\#24$ &
    & Perform actions on user behalf &
    \rc & \gc & \rc & \rc \\[2pt]
    \cline{2-8}
    & $\#25$ &
    & Associate arbitrary accessibility-related information to a view &
    \rc & \gc & \rc & \rc \\[2pt]
    \hline

    \end{tabular}
\end{table*}
Table~\ref{tab:concepts} also indicates which APIs implement \emph{basic} screen reader functionalities.
To classify \emph{basic} functionalities
we took into account the accessibility principles and best practices presented in introductory accessibility documentation by Apple and Google~\cite{AppleAccDoc,GoogleAccDoc}.
Then, we considered the APIs mentioned in these resources as the \textit{basic} ones. 


The identified APIs are organized into five categories, described in details in the following.

\subsection{Accessibility focus}
A basic principle in mobile screen-readers interaction is that the user can select a view, hear its description and then possibly activate it.
When a view is selected, we say it receives the \textit{accessibility focus}; at most one view can have the accessibility focus at the same time.
Not all views should receive the accessibility focus. For example, an image should not receive the accessibility focus if it purpose is solely to improve the quality of the visual presentation. Screen reader API $\#1$ allows the developer to specify which views should receive the accessibility focus\footnote{We use the notation $\#n$ to refer to the API functionality with id $n$ in Table~\ref{tab:concepts}.}.

There are two possible forms of interaction to move the accessibility focus among views: with the \emph{sliding exploration} the user slides the finger on the screen and a view gets the accessibility focus once touched. Instead, with the \emph{sequential exploration} the user enters right/left swipe gestures that move the focus sequentially towards the next/previous view.
This latter interaction requires the views to be ordered with a numerical index called \textit{accessibility focus order}. Screen-reader API $\#2$ allows the developer to personalize the accessibility-focus order.
When not explicitly specified by the developer, the accessibility focus order is automatically derived from views' position on the screen.

There are situations in which a developer needs to programmatically assign the accessibility focus to a view (API $\#3$), for example when the keyboard is shown, and the developer wants to set the accessibility focus to the first key. Similarly, the developer might want to run arbitrary code when a view receives or loses the accessibility focus (API $\#4$), for example by playing an audio cue when an icon is selected (an interaction technique commonly known as audio-icons~\cite{gaver1986auditory}). Finally, the developer may need to determine which view has the accessibility focus (API $\#5$).

Table~\ref{tab:concepts} shows that \textit{iOS} and \textit{Android} expose functionally equivalent APIs for all these functionalities.

\subsection{Text to announce}
When a view receives the accessibility focus, the screen-reader needs to read aloud the textual description associated to it.
This textual description, which we call \emph{text-to-announce}, can be defined by the developer.

\textit{VoiceOver} and \textit{TalkBack} adopt two different approaches to allow the developer to personalize the text-to-announce.
In \textit{\textit{iOS}} the developer can specify, for each view, some attributes that \textit{VoiceOver} uses to compute the text-to-announce. For example, \textit{accessibilityLabel} is a text that identifies the view and that forms the first part of the text-to-announce; similarly \textit{accessibilityHint} specifies the effects of the interaction and is one of the last pieces of information forming the text-to-announce.
Similarly to \textit{\textit{iOS}}, \textit{\textit{Android}} allows the developer to specify a field, called \textit{contentDescription}, which is a textual attribute that can be read as part of the text-to-announce.
However, differently from \textit{\textit{iOS}}, in \textit{\textit{Android}} the developer can also programmatically specify the text-to-announce. So there are cases in which  \textit{contentDescription}, although specified, is ignored. For example, a view containing editable text announces the \textit{contentDescription} when the contained text is empty, and otherwise it announces the contained text ignoring the \textit{contentDescription}.
This difference is captured by APIs $\#6$, which is exposed by \textit{iOS} and \textit{Android}, and $\#7$, exposed by \textit{Android} only.

There are other two APIs differentiating the two platforms and that are exposed by \textit{Android} and not by \textit{iOS}. The first API ($\#8)$ allows the developer to specify that one view describes another one, as for example in the case of a label describing an editable text field. The second difference is in API $\#9$, to specify that a view should be announced also when the user is not interacting with it (e.g., an ``incorrect password'' label that is shown during a login procedure).

Finally, both native platforms allow the developer to specify the language of the text-to-announce ($\#10$).

\subsection{Explicit text-to-speech}
The text-to-announce is read aloud by a text-to-speech (TTS) software. The same TTS can also be used by the developer to programmatically read a text aloud. This is useful, for example, to inform the user that a view has moved.

\textit{iOS} and \textit{Android} provide two ways to access the TTS (APIs $\#11$ and $\#14$): as an accessibility function (which we call \emph{screen reader TTS}) and as a functionality unrelated to accessibility (\emph{non-screen reader TTS}). In the former case the text will be read aloud only when the screen reader is running, while in the latter case the text is also read aloud when the screen reader is off.
In both cases, it is often necessary to synchronize the TTS output with other audio output (in many situations the developer does not want two or more audio outputs to overlap).
This is possible for both types of TTS on \textit{iOS} (APIs $\#12$ and $\#15$), and with the non-screen reader TTS only on \textit{Android} (API $\#15$).

\textit{iOS} and \textit{Android} also expose API $\#17$ to personalize how the TTS reads the text aloud, for example changing the language, the speed, etc. With \textit{iOS} this can be personalized also for screen reader TTS (API $\#13$).

\subsection{Accessibility tree}
In \textit{iOS} and \textit{Android} a view can contain other views. In this case we say that the former is a \emph{container}.
The \emph{contained-in} relationship among views defines a structure, which we call \textit{view tree}: each view can be either a container or a leaf node (i.e., a non-container view) and there is a single container that acts as a root (it represents the entire screen).

There is also a similar tree, which we call \emph{accessibility tree}, that represents the view tree from the point of view of a screen reader. We call \textit{accessible element} the nodes in the accessibility tree and, as we detail in the following, each accessible element can correspond to a view, a group of views or a portion of a view.

By default, the views and accessibility trees coincide.
However, there are two cases in which it is convenient to differentiate them.
One case is when there are multiple views that should be presented as a single accessibility element. Consider for example a list in which each cell is composed of many views: by aggregating these views in a single accessible element the user can find list elements more easily using sliding exploration and can scroll it faster using sequential exploration.
This behaviour can be implemented in \textit{iOS}, but the developer should specify the code to aggregate multiple views in a single accessible element\footnote{See, for example, the \emph{Group Labels} section in the Apple \emph{Delivering an Exceptional Accessibility Experience} sample code \url{https://apple.co/3cT6k6l}}. Instead, in \textit{Android} a single API ($\#18$) can be used to conveniently implement this.

The other case in which it is convenient to differentiate the view and accessibility trees is when a single view contains  multiple pieces of information that should be provided individually to the user. Consider for example an image that can be tapped in various positions to trigger different actions. In this case the image view can be divided into multiple accessible elements, each one representing different portions of the image and triggering different actions.
Both \textit{iOS} and \textit{Android} expose APIs to implement this (see $\#19$).

\textit{iOS} and \textit{Android} also provide APIs to navigate the accessibility tree and in particular to get the parent accessibility element (API $\#20$).


 
\subsection{Miscellaneous}
\textit{iOS} and \textit{Android} allow the developer to query whether the screen reader is active (API $\#21$). This is useful, for example, when the developer wants to personalize the app behaviour for screen reader users or to remotely log screen reader usage data so that app accessibility can be evaluated and eventually improved. 

In this category \textit{Android} exposes three APIs that are not available in \textit{iOS}. The first ($\#22$) allows the developer to specify that a view is a header, which helps the user to navigate in the screen. The second ($\#24$) can be used to programmatically trigger view events (e.g., activate a button); this can be used to support interaction automation within the app.
The last API ($\#25$) allows the developer to associate arbitrary information related to accessibility with any view. This information can be used, for example, to personalize the text-to-speech with code specifically designed for this purpose.

One set of APIs ($\#23$) in this category is instead exposed by \textit{iOS} and not by \textit{Android}. It allows to define actions that can be triggered only when the screen reader is active. Among others, these APIs allow the developer to define actions that are specific to \textit{VoiceOver}, like the rotor actions and the \textit{magic tap}\footnote{\emph{Magic tap} is a gesture (double tap with two fingers) that can be entered anywhere in the app and that toggles the most important state of the app (e.g., pause/play the music)}.





\section{CPDF Screen reader API availability}
The last two columns of Table~\ref{tab:concepts} report whether each functionality is available in \textit{React Native} or \textit{Xamarin}, respectively.
We denote that the functionality is available ($\gc{}$ symbol) in two cases: (i) when both \textit{iOS} and \textit{Android} expose a functionally equivalent API that is wrapped by the CPDF producing the same behaviour on both platforms, and (ii) when an API is available in either \textit{iOS} or \textit{Android} and the CPDF wraps the API for that native platform, producing the expected behaviour in it and not in the other native platform. 
We indicate that a API is not exposed by a CPDF with the \rc{} symbol, while we use the \ec{} symbol to denote that the API is exposed by the CPDFs, but with a possibly different behaviour in the two native platforms.

%

In Table~\ref{tab:concepts} we can observe that in most cases ($14$ out of $25$), the same functionality is available on both \textit{iOS} and \textit{Android}.
In all of these cases it would be possible for CPDF to wrap native APIs into a single cross-platform API, so that the developers can easily access the native APIs for that functionality.
In practice, however, this is not the case in both the examined CPDF and, in particular in \textit{Xamarin}. Indeed, out of the $14$ functionalities shared by \textit{iOS} and \textit{Android}, only $8$ and $5$ are wrapped into \textit{React Native} and \textit{Xamarin} APIs, respectively.
For the APIs that are exposed by only one native platform, it could still be possible for CPDF to wrap the API for that platform. Again, this is only rarely the case. Indeed, out of $11$ API exposed by \textit{iOS} or \textit{Android} (but not both), only $5$ and $2$ are wrapped by \textit{React Native} and \textit{Xamarin}, respectively.
The situation is only slightly better considering basic functionalities only. Indeed, out of $8$ basic functionalities, only $6$ and $5$ are available in \textit{React Native} and \textit{Xamarin}, respectively.
In the following of this section we detail API availability for each category.

\subsection{Accessibility focus}
For this category, \textit{iOS} and \textit{Android} expose functionally equivalent APIs. 
\emph{React Native} and \emph{Xamarin}, instead, do not expose some APIs, including $\#2$ (\textit{React Native}), $\#3$ (\textit{Xamarin}), $\#4$ and $\#5$ (both).
Interestingly, the functionality to specify the accessibility focus order ($\#2$) which is not exposed by \textit{React Native} is a basic accessibility functionality.

\subsection{Text-to-announce}
As explained above, \textit{iOS} and \textit{Android} expose a different specification of the text-to-announce (API $\#6$).
However, this difference is not well captured by the APIs exposed by  \textit{React Native} and \textit{Xamarin}. For example, both CPDFs allow the developer to specify a field (\textit{accessibilityLabel} in \textit{ReactNative} and \textit{AutomationProperties.SetName} in \textit{Xamarin}) that wraps the field \textit{accessibilityLabel} for \textit{iOS} and \textit{contentDescription} for \textit{Android}. However these two fields produce a similar but not identical behaviour, and this can lead to unexpected results when developing with \textit{ReactNative} or \textit{Xamarin}.

Another problem arises in \textit{Xamarin} again for API $\#6$: text fields (\textit{e.g.},  \textit{Label} or \textit{Entry}) erroneously wrap the \textit{AutomationProperties.SetName} field to the \textit{Android} \textit{hint} attribute, which is not designed for accessibility. The results into two possibly unpredictable behaviours. First the \textit{AutomationProperties.SetName} attribute is correctly read as part of the text-to-announce only when the text field is empty. Second, when the text field is empty, the hint text is shown, which is unexpected for a field that should specify the text-to-announce. 

Both \textit{React Native} and \textit{Xamarin} do not provide support for APIs $\#7$ and $\#10$. Instead, \textit{React Native} supports API $\#9$ while \textit{Xamarin} expose API $\#8$, which is a basic accessibility functionality. 


\subsection{Explicit TTS.}
\textit{React Native} exposes most APIs in this category, supporting both screen reader TTS and non-screeen reader TTS. The only APIs that are not exposed are $\#13$ and $\#16$.

Instead, \textit{Xamarin} does not support any API related to screen reader TTS and only supports non-screen reader TTS (APIs $\#14$ and $\#17$). Note, in particular, that \textit{Xamarin} does not expose API $\#11$, which is a basic accessibility functionality.

\subsection{Accessibility tree}

While \textit{iOS} and \textit{Android} present functionally equivalent APIs (with the only exception of API $\#18$ that, however, is not strictly necessary), \textit{React Native} and \textit{Xamarin} only expose API $\#18$.




\subsection{Miscellaneous}
\textit{Xamarin} does not expose any APIs in this category, including $\#22$ and $\#23$ that are basic accessibility functionalities.
Instead, \textit{React Native} exposes API $\#21$, $\#22$
and $\#23$, including those to manage \textit{iOS} rotor.
However, no support is provided for API $\#24$ and $\#25$.




\section{Implementation}
In this section we describe the sample code that we distribute online and we overview the main technical challenges arising when accessing the native APIs that are not wrapped by \textit{React Native} and \textit{Xamarin}.

\subsection{Sample apps}
We selected a subset of the APIs from Table~\ref{tab:concepts} (including all basic functionalities) and for each of them we designed a sample app, which we then implemented for the considered platforms (\textit{iOS}, \textit{Android}, \textit{React Native} and \textit{Xamarin}).
The source code is available online\footref{fn:link}.

\begin{figure*}[t]
	\centering
	\subfloat[API $\#1$ (Native \textit{iOS})]{
		\label{fig:api1}
		\includegraphics[height=.4\textwidth]{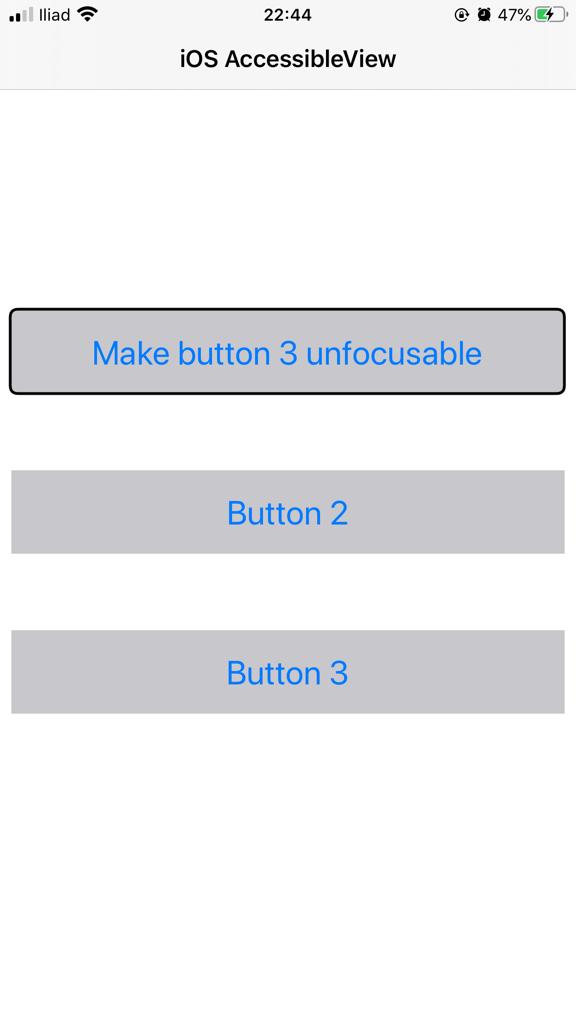}
	}\hfill
	\subfloat[API $\#2$ (native \textit{Android})]{
		\label{fig:api2}
		\includegraphics[height=.4\textwidth]{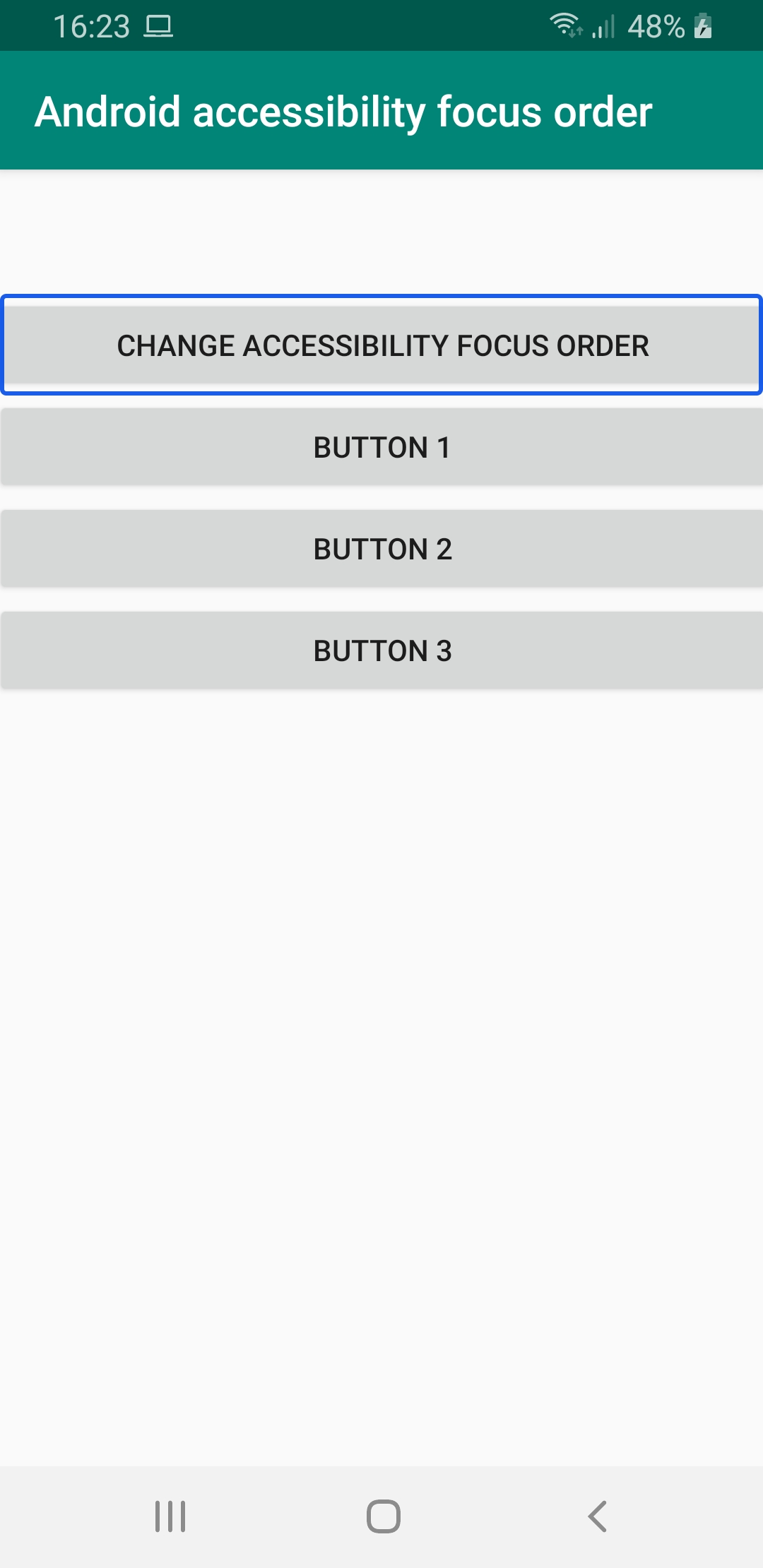}
	}\hfill
	\subfloat[API $\#3$ (\textit{React Native} deployed on \textit{Android})]{
		\label{fig:api3}
		\includegraphics[height=.4\textwidth]{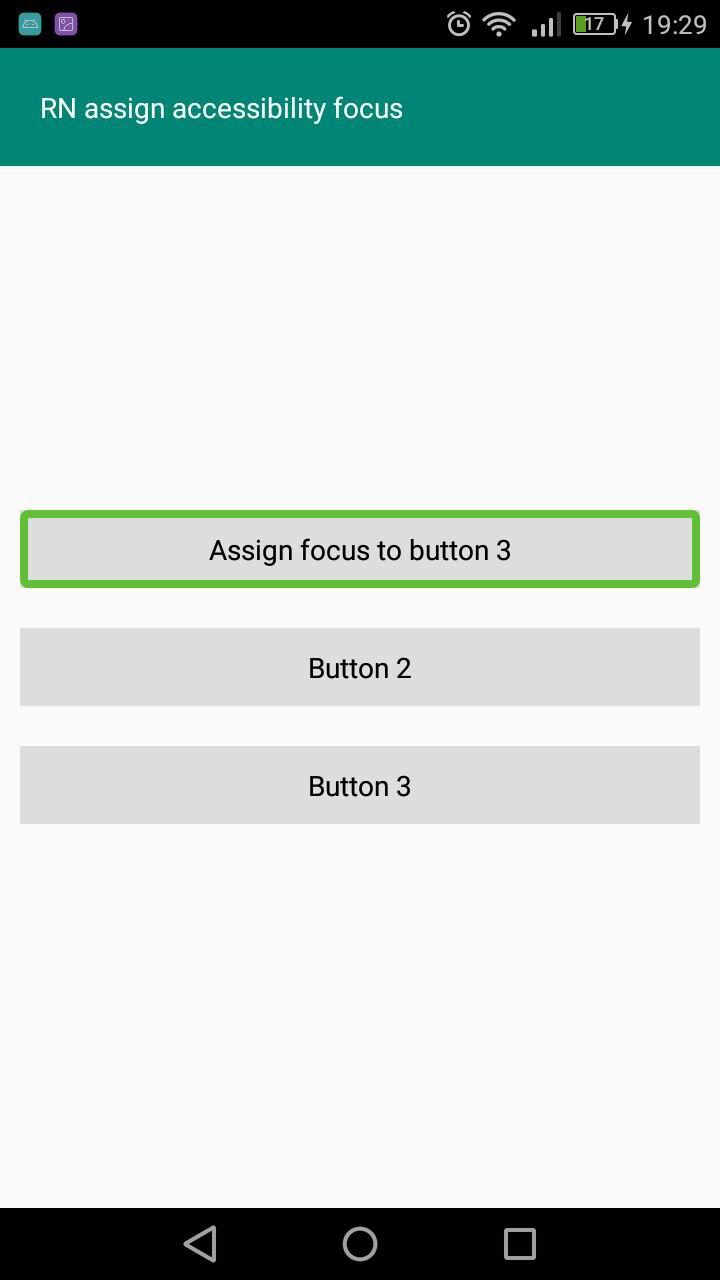}
	}\hfill
	\subfloat[API  $\#4$ (\textit{Xamarin} deployed in \textit{iOS})]{
		\label{fig:api4}
		\includegraphics[height=.4\textwidth]{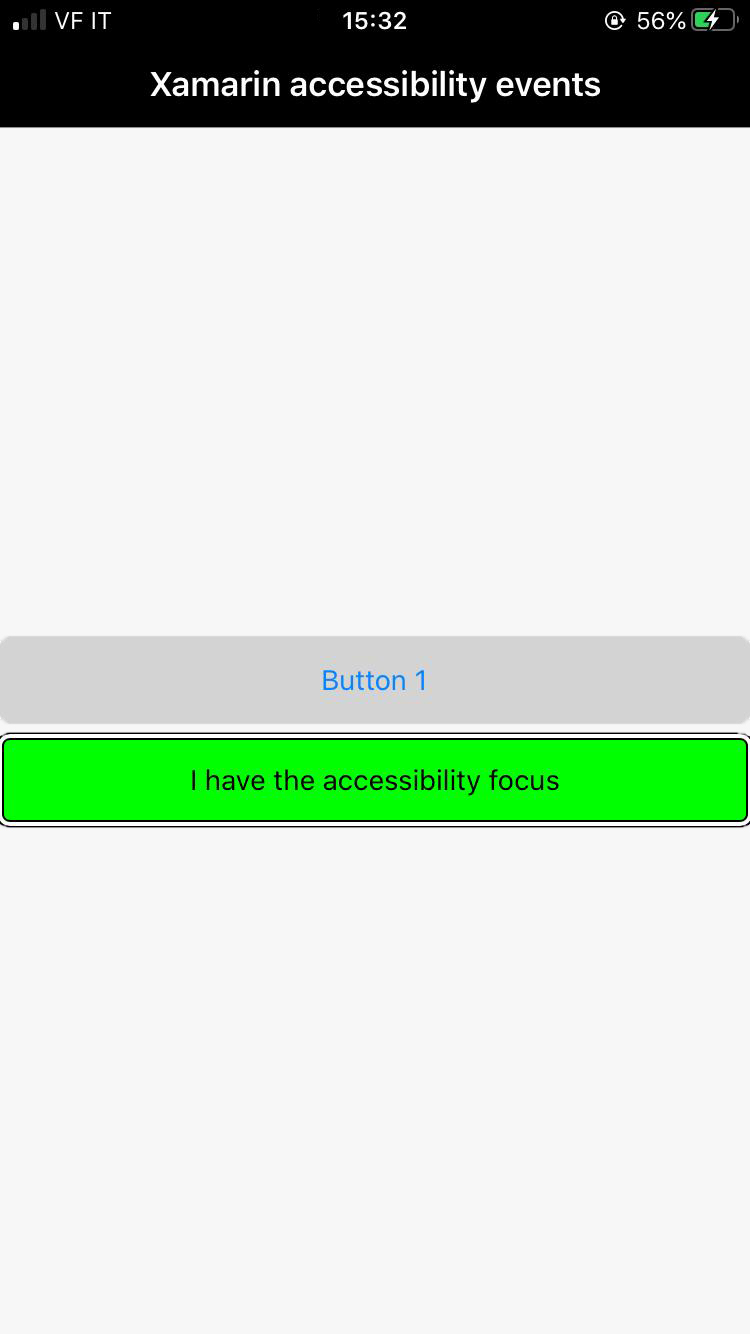}
	}
	\caption{Examples of the samples applications}
	\label{fig:sampleApps}
\end{figure*}

Figure~\ref{fig:sampleApps} shows some examples. In particular, Figure~\ref{fig:api1} shows the sample app  for API $\#1$ implemented in native \textit{iOS}: the app shows three buttons and initially they can all receive the accessibility focus; activating the first button results in the third button to become not focusable any more.
Figure~\ref{fig:api2} shows the sample app for API $\#2$ implemented in native \textit{Android}: the app shows four buttons. Initially the accessibility focus order is the same as the visual order but when the first button is activated the accessibility focus order is changed. 
The third example (see Figure~\ref{fig:api3}) is the sample app for API $\#3$ implemented in \textit{React Native} and deployed to \textit{Android}. The app shows three buttons: when the first one is activated, the accessibility focus is assigned to Button 3.  
Figure~\ref{fig:api4} shows the sample app for API $\#4$ implemented in \textit{Xamarin} and deployed to \textit{iOS}. The app shows two buttons: when the second one receives the accessibility focus, it changes its color and label.

As detailed in Table~\ref{tab:implemented}, we implemented each sample app in native code (\textit{iOS} and \textit{Android}) and in the two considered CPDF, each being deployable both in \textit{iOS} and \textit{Android}.
In few cases the implementation is missing (denoted with an empty cell in Table~\ref{tab:implemented}) because it is not supported by the given native platform and it is not possible to implement it using a combination of other APIs (\textit{e.g}., the rotor is not available in \textit{Android}). 
We classify the sample apps depending on how the accessibility functionality is implemented:
\begin{itemize}
    \item \emph{Direct API use}: the accessibility functionality is implemented directly using the API exposed by the platform (\direct{} symbol);
    \item \emph{Semi-direct API use}: the accessibility functionality is not exposed by the platform API, but a similar behaviour can still be achieved by using a combination of APIs exposed by the platform (\semidirect{} symbol);
    \item \emph{Native API use} (for applications developed in CPDF only): APIs are not exposed by the platform and the implementation requires to access native APIs (\indirect{} symbol).
\end{itemize}

For example, functionality $\#1$ can be implemented with direct API use in all four platforms, as they all expose API for this.
Instead, \textit{iOS} does not expose an API for functionality $\#8$, but it still can be implemented with semi-direct API use, because it can be implemented using a combination of API $\#1$ and $\#6$ (both available in \textit{iOS}).
An example of native API use is functionality $\#4$ in \textit{React Native}: no API is available and it is not possible to implement the functionality using a combination other APIs available in the platform, hence it is necessary to use the native APIs available in \textit{iOS} and \textit{Android}.




\begin{table}[h!]
    \fontsize{7.5}{14}\selectfont
    \centering
    \caption{Sample applications and how they are implemented (\direct{} = direct, \semidirect{} = semi-direct, \indirect{} = native)}
    \label{tab:implemented}
    \begin{tabular}{|c||c|c||c|c||c|c||}
    \hline
     & \multicolumn{2}{c||}{Native} & \multicolumn{2}{c||}{\textit{React Native}} & 
    \multicolumn{2}{c||}{\textit{Xamarin}} \\ 
        \hline

    \textbf{ID} &
    \textbf{{iOS}} &
    \textbf{{Android}} &
    \textbf{{iOS}} &
    \textbf{{Android}} &
    \textbf{{iOS}} &
    \textbf{{Android}}
\\ 
    \hline
    $\#1$ &
    \direct{} & \direct{} & \direct{} & \direct{} & \direct{} & \direct{} \\
    \hline
    $\#2$ &
    \direct{} & \direct{} & \indirect{} & \indirect{} & \direct{} & \direct{} \\
    \hline
    $\#3$ &
    \direct{} & \direct{} & \direct{} & \direct{} & \indirect{} & \indirect{}\\
    \hline
    $\#4$ &
    \direct{} & \direct{} & \indirect{} & \indirect{} & \indirect{} & \indirect{} \\
    \hline
    $\#6$ &
    \direct{} & \direct{} & \direct{} & \direct{} & \direct{} & \direct{} \\
    \hline
    $\#8$ &
    \semidirect{} & \direct{} & \semidirect{} & \semidirect{} & \semidirect{} & \direct{} \\
    \hline
    $\#11$ &
    \direct{} & \direct{} & \direct{} & \direct{} & \indirect{} & \indirect{} \\
    \hline
    $\#18$ &
    \semidirect{} & \direct{} & \direct{} & \direct{} & \semidirect{} & \direct{} \\
    \hline
    $\#22$ &
     & \direct{} & & \direct{} & & \indirect{} \\
    \hline
    $\#23$ &
    \direct{} &  & \direct{} & & \indirect{} & \\
    \hline
    
    \end{tabular}
\end{table}
From Table~\ref{tab:implemented} we can observe that implementing a functionality in CPDF is not possible if the API is not available in the target native platform. For example, consider API $\#22$: is not available in \textit{iOS}, hence cannot be implemented in \textit{React Native} or \textit{Xamarin} when deploying to \textit{iOS}.
Similarly, the implementation in CPDF is at least as difficult as in native code (native API use is more difficult than semi-direct APi use that is more difficult than direct API use). There is only one exception for API $\#18$: this functionality can be implemented in \textit{iOS} with semi-direct API use. However, \textit{React Native} exposes an API that automatically implements the same behaviour of the \textit{Android} call also in \textit{iOS}. For this reason we report it as a direct API use. 

\subsection{Implementation challenges and effort}
Considering the \textit{Xamarin} implementation, as described in Table~\ref{tab:implemented}, we developed six sample apps using native API use: $\#3$, $\#4$, $\#8$, $\#11$, $\#22$, $\#23$.
All of them have been implemented using \textit{dependency services} except for sample app for API $\#4$, for which it was necessary to use \textit{custom renderers}.
In both cases (\textit{dependency services} and \textit{custom renderers}) one challenge for the developer is to be aware of the \textit{Android} and \textit{iOS} native APIs, their (sometimes subtle) differences, and how they are wrapped into the $C\#$ platform-specific APIs. This typically requires the developer to deal with the documentation for native \textit{iOS}, native \textit{Android}, \textit{Xamarin.\textit{iOS}} and \textit{Xamarin.\textit{Android}}.
An additional challenge with \textit{custom renderers} is that the developer also needs to know how \textit{Xamarin Forms} \textit{renderers} work, and to be aware of all the classes involved in this part of the code.

To give an idea of the effort required to create accessible application in \textit{Xamarin}, we report, as an example, our own experience with the sample apps, created by a novice \textit{Xamarin} developer (a co-author of this paper) supported by two professional \textit{Xamarin} developers, with $1$ and $3$ years of experience in \textit{Xamarin} development, respectively, and a team of experts in mobile accessibility on native platforms. Each sample app with direct API use required about one hour of development time, requiring limited support.
The development time was similar for the applications based on semi-direct API use, but in some cases some support was needed by the team of experts in mobile accessibility to define how to combine exposed API to implement the expected behaviour. 
Instead, sample apps with native API use required a much higher effort. In particular, those created using \textit{dependency services} required about $2$ hours of development each, and the intervention of the professional developers. Sample app for API $\#4$, which required to use \textit{custom renderers} required several weeks of investigation and despite the involvement of both supervising developers we were still unable to solve the problem. 
Eventually, we succeeded after receiving help from the official \textit{Xamarin} support provided by a \textit{Microsoft} developer. 

Considering the implementation in \textit{React Native}, sample apps for API $\#2$ and $\#4$ were developed with native API use. This required, for each sample app, to develop a \emph{native component}, which is composed of a few classes in native \textit{Android} and \textit{iOS}, hence written in \textit{Java} and \textit{ObjectiveC}, respectively.
Each native component implements the accessibility functionalities by accessing native APIs and exposing them to the remaining of the \textit{React Native} code.
Similarly to what happens with \textit{Xamarin}, the developer needs to know the native accessibility APIs. However, in the case on \textit{React Native} the developer also needs to know how to code in native programming languages (\textit{Java} and \textit{ObjectivC}) and how the \textit{React Native} native libraries work, and this is particularly challenging due to the fact that they are poorly documented. 

Also in this case we report out experience with the development of the sample apps, created by one of the co-authors of this paper who is a professional developer with more than $3$ years of experience in development with \textit{React Native} and no previous experience in the development of native components.
On average, he required less than one hour for each of the sample apps requiring direct or semi-direct API use.
Instead, he required more than $10$ days to complete sample app for API $\#4$ on \textit{Android} and he also needed to be supported by another experienced \textit{iOS} developer to complete the \textit{iOS} native component.

\section{Conclusions and future work}
It is well known that many applications, including those for mobile devices, have accessibility issues~\cite{ross2017epidemiology}.
In some cases, this is due to a lack of knowledge by the developers, who are not aware of the needs of people with disabilities, or do not know how to address these needs with the existing technology.

This paper uncovers a different problem: while native development platforms offer a broad range of APIs to develop accessible applications, many of these APIs are not wrapped by CPDF. This means, for example, that a developer who is aware of the needs of people with visual impairments and who knows the native APIs to guarantee accessibility, will still struggle to implement accessibility functionalities in \textit{React Native} and \textit{Xamarin}. As we show in this paper, the developer can still implement these functionalities, but this require a higher development effort; in a commercial project this implies higher costs, hence it lowers the chances that the functionality is actually implemented.

Another relevant contribution of this paper is that it lists the accessibility functionalities exposed by \textit{iOS} and \textit{Android}. We do not have formal guarantees that this list is complete, but it represents a starting point for a discussion in the scientific and technical community. We believe that the list of accessibility APIs represents a valid technical reference documentation. It can be used by developers interested in creating accessible applications, including in particular mobile assistive technologies.
Also, developers of new CPDFs and mobile OSs can refer to this paper to quickly understand which accessibility functions they need to implement.

This paper analyses an involved technical scenario and identifies a new relevant problem, hence paving the way for a number of follow-up research directions and technical interventions.
On one side, this paper suggests that CPDF can contribute to the development of applications presenting accessibility issues. This hypothesis can be verified in the future, for example with large scale studies, like the one conducted by Ross et al.~\cite{ross2017epidemiology}.
The same methodology presented in this paper can be adapted to analyze the accessibility of other systems, for example traditional devices (desktop computers) and pervasive ones, or to address the accessibility to mobile devices by people with other forms of disabilities, like low vision.
On the technical side, this paper identifies a intervention checklist for the development community to implement new functionalities in \textit{React Native} and \textit{Xamarin}. Both platforms are open source and hence it is possible to contribute in the development of new functionalities, which we intend to do.
In the future we also intend to consider other CPDFs, including  \textit{Flutter}.



%


\bibliographystyle{SIGCHI-Reference-Format}
\bibliography{main}

\section{Supplementary Materials}
Tables~\ref{tab:nativescreenapi} and \ref{tab:crossplatformscreenapi} report how APIs are implemented in native and CPDF platforms respectively. Some of the APIs are available both as dynamic methods that can be called in the code and as static properties that can be set in XML view representations. In Tables~\ref{tab:nativescreenapi} and \ref{tab:crossplatformscreenapi} we report the dynamic methods and indicate with the \sm{} symbol when a static property is also available. For \textit{React Native} we instead report the XML attributes (no dynamic code is available in many cases).


\begin{table*}[h!]
    \fontsize{7.5}{11}\selectfont
    \centering
    \caption{Screen reader APIs for iOS and Android.  Notation: \rc{} = non available, \sm{} = also available as a static attribute}
    \label{tab:nativescreenapi}
    \begin{tabular}{|p{0.4cm}|p{0.3cm}|p{7cm}|p{7cm}|}
    \hline
    \rotatebox[origin=c]{90}{\textbf{Category}} &
    \textbf{ID} &                                  
    \textbf{\centering iOS} &
    \textbf{\centering Android} \\ 
    \hline
    \hline
    
    \multirow{7}{*}{\rotatebox[origin=c]{90}{Acc. focus}} &
    $\#1$ &
    \begin{tabular}[c]{@{}l@{}}
        \code{UIView.isAccessibilityElement} \sm{}\\ \code{UIAccessibility.accessibilityViewIsModal} \\
        \code{UIAccessibility.accessibilityElementsHidden}
    \end{tabular} &
    \begin{tabular}[c]{@{}l@{}}
        \code{View.setImportantForAccessibility} \sm{}\\
    \end{tabular} \\
    
    \cline{2-4}
    & $\#2$ &
    \begin{tabular}[c]{@{}l@{}}
    \code{UIAccessibilityContainer.UIView.accessibilityElements} \\
    \code{UIAccessibility.shouldGroupAccessibilityChildren}
    \end{tabular}
    &
    \code{View.setAccessibilityTraversalBefore} (or \code{After}) \sm{}\\
    \cline{2-4}
    & $\#3$ &
    \code{UIAccessibility.post(notification:.layoutChanged, argument: destinationView)} &
    \code{View.sendAccessibilityEvent(AccessibilityEvent.WINDOWS\_CHANGE\_ACCESSIBILITY\_FOCUSED)} \\
    \cline{2-4}
    & $\#4$ &
    \code{UIAccessibilityFocus.accessibilityElementDidBecomeFocused} (or \code{DidLoseFocus}) &
    \code{View.AccessibilityDelegate.onInitializeAccessibilityEvent} \\
    \cline{2-4}
    & $\#5$ &
    \code{View.accessibilityElementIsFocused} &
    \code{AccessibilityNodeProvider.findFocus} \\
    \hline
    \hline
    
    \multirow{5}{*}{\rotatebox[origin=c]{90}{Text to ann.}} &
    $\#6$ &
    \begin{tabular}[c]{@{}l@{}}
       \code{UIAccessibility.accessibilityLabel} \sm{}\\
        \code{UIAccessibility.accessibilityValue}\\
        \code{UIAccessibility.accessibilityHint} \sm{}\\
        \code{UIAccessibility.accessibilityTraits} \sm{}\\
        \code{UIAccessibility.accessibilityRespondsToUserInteraction}\sm{}
    \end{tabular} &
    \begin{tabular}[c]{@{}l@{}}
        \code{View.setContentDescription} \sm{}\\
        \code{View.getAccessibilityClassName}
    \end{tabular} \\
    \cline{2-4}
    & $\#7$ &
    \rc &
    \begin{tabular}[c]{@{}l@{}}
        \code{View.AccessibilityDelegate}\\
        \code{View.onPopulateAccessibilityEvent}\\
        \code{View.onInitializeAccessibilityNodeInfo}
    \end{tabular} \\
    \cline{2-4}
    & $\#8$ &
    \rc &
    \code{View.setLabelFor} \sm{}\\
    \cline{2-4}
    & $\#9$ &
    \rc &
    \code{View.setAccessibilityLiveRegion} \sm{}\\
    \cline{2-4}
    & $\#10$ &
    \code{UIAccessibility.accessibilityLanguage} &
    see \code{SpannableStringBuilder} \\
    \hline
    \hline
    
    \multirow{7}{*}{\rotatebox[origin=c]{90}{Explicit TTS}} &
    $\#11$ &
    \begin{tabular}[c]{@{}l@{}}
        \code{UIAccessibility.post(notification: .announcement,}\\
        \code{argument: stringToRead)}\\
        or \code{.pageScrolled}\\
        or \code{.screenChanged}
    \end{tabular} &
    \code{View.announceForAccessibility} \\
    \cline{2-4}
    & $\#12$ &
    \code{UIAccessibility.announcementDidFinishNotification} &
    \rc \\
    \cline{2-4}
    & $\#13$ &
    \code{NSAttributedString.Key} &
    \rc \\
    \cline{2-4}
    & $\#14$ &
    \code{AVSpeechSynthesizer.speak} &
    \code{TextToSpeech.speak} \\
    \cline{2-4}
    & $\#15$ &
    \code{AVSpeechSynthesizer.isSpeaking} &
    \code{TextToSpeech.isSpeaking} \\
    \cline{2-4}
    & $\#16$ &
    \code{AVSpeechSynthesizer.pauseSpeaking} &
    \rc{} \\
    \cline{2-4}
    & $\#17$ &
    \begin{tabular}[c]{@{}l@{}}
        \code{AVSpeechUtterance.rate}\\
        or \code{voice}\\
        or \code{pitchMultiplier}
    \end{tabular} &
    \begin{tabular}[c]{@{}l@{}}
            \code{TextToSpeech.setPitch}\\
            or \code{setSpeechRate}\\
            or \code{setVoice}
    \end{tabular} \\
    \hline
    \hline
    
    \multirow{5}{*}[+2.5ex]{\rotatebox[origin=c]{90}{A. T.}} &
    $\#18$ &
    \begin{tabular}[c]{@{}l@{}}
           \rc
    \end{tabular} &
    \code{View.focusable} (was: \code{screenReaderFocusable}) \sm{}\\
    \cline{2-4}
    & $\#19$ &
    \begin{tabular}[c]{@{}l@{}}
       \code{UIAccessibilityContainer}\\ \code{UIAccessibilityElement} 
    \end{tabular} &
    \begin{tabular}[c]{@{}l@{}}
       \code{AccessibiliyNodeProvider} \\ \code{AccessibilityNodeInfo} 
    \end{tabular} \\
    
    \cline{2-4}
    & $\#20$ &
    \code{UIAccessibilityElement.accessibilityContainer} &
    \code{View.getParentForAccessibility} \\
    \hline
    \hline
    
    \multirow{4}{*}[-6.4ex]{\rotatebox[origin=c]{90}{Misc.}} &
    $\#21$ &
    \begin{tabular}[c]{@{}l@{}}
    \code{UIAccessibility.isVoiceOverRunning}\\
     \code{UIAccessibility.voiceOverStatusDidChangeNotification}
    \end{tabular} &
    \code{AccessibilityManager.isEnabled} \\
    \cline{2-4}
    & $\#22$ &
    \rc &
    \begin{tabular}[c]{@{}l@{}}
        \code{View.setAccessibilityHeading} \\
        \code{View.setAccessibilityPaneTitle} 
    \end{tabular} \\
    \cline{2-4}
    & $\#23$ &
    \begin{tabular}[c]{@{}l@{}}
        \code{UIAccessibility.accessibilityCustomRotor}\\ \code{UIAccessibilityCustomAction}\\
        In UIAccessibility among others: \\ \code{UIAccessibility.accessibilityPerformMagicTap} \\
        or \code{accessibilityAction} \\
        or \code{accessibilityIncrement} \\
        or \code{{accessibilityDecrement}}
    \end{tabular} &
    \rc \\
    \cline{2-4}
    & $\#24$ &
    \rc &
    \code{AccessibilityNodeProvider.performAction} \\
    \cline{2-4}
    & $\#25$ &
    \rc &
    \code{AccessibilityNodeProvider.addExtraDataToAccessibilityNodeInfo} \\
    \hline

    \end{tabular}
\end{table*}


\begin{table*}[h!]
    \fontsize{7.5}{11}\selectfont
    \centering
    \caption{Screen reader APIs in \textit{React Native} and \textit{Xamarin} (\rc = non available, \sm{} = also available as a static attribute)}
    \label{tab:crossplatformscreenapi}
    \small
    \begin{tabular}{|p{0.4cm}|p{0.3cm}|p{7cm}|p{7cm}|}
    \hline
    \rotatebox[origin=c]{90}{\textbf{Category}} &
    \textbf{ID} &                                  
    \textbf{\centering React Native} &
    \textbf{\centering Xamarin} \\ 
    \hline
    \hline
    
    \multirow{7}{*}[3.5ex]{\centering\rotatebox[origin=c]{90}{Acc. focus}} &
    $\#1$ &
    \begin{tabular}[c]{@{}l@{}}
        \code{<View} \code{accessible={true}/>} \\
        \code{<View} \code{accessibilityViewIsModal={true} />} (only iOS)\\
        \code{accessibilityElementsHidden} (iOS) \\
        \code{importantForAccessibility="no-hide-descendants"} (Android)
    \end{tabular} &
    \code{AutomationProperties.SetIsInAccessibleTree} \sm{} \\
    \cline{2-4}
    & $\#2$ &
    \rc &
    \code{View.TabIndex} \sm{}\\
    \cline{2-4}
    & $\#3$ &
    \code{AccessibilityInfo.setAccessibilityFocus} &
    \rc \\
    \hline
    \hline
    
    \multirow{5}{*}[4.5ex]{\rotatebox[origin=c]{90}{Text to ann.}} &
    $\#6$ &
    \begin{tabular}[c]{@{}l@{}}
        \code{<TouchableOpacity} \code{accessible={true}} \\
        \code{accessibilityLabel= mystring} \\
        \code{accessibilityHint= mystring} \\
        \code{accessibilityState} \\
        \code{accessibilityValue} \\
        \code{accessibilityRole />}
    \end{tabular} &
    \begin{tabular}[c]{@{}l@{}}
        \code{AutomationProperties.SetName} \sm{}\\
        \code{AutomationProperties.SetHelpText} \sm{}\\
    \end{tabular} \\
    \cline{2-4}
    & $\#8$ &
    \rc &
    \code{AutomationProperties.SetLabeledBy} \sm{}\\
    \cline{2-4}
    & $\#9$ &
    \code{accessibilityLiveRegion} (only Android) &
    \rc \\
    \cline{2-4}
    \hline
    \hline
    
    \multirow{6}{*}[-5.5ex]{\rotatebox[origin=c]{90}{Explicit TTS}} &
    $\#11$ &
    \code{AccessibilityInfo.announceForAccessibility} &
    \rc \\
    \cline{2-4}
    & $\#12$ &
    \code{AccessibilityInfo.setEventListener("announcementFinished", handler)} (only iOS) &
    \rc \\
    \cline{2-4}
    & $\#14$ &
    \begin{tabular}[c]{@{}l@{}}
        \code{Tts.speak(mystring, params)} (using react-native-tts) \\
        \code{Speech.speak} (using Expo SDK)
    \end{tabular} &
    \code{TextToSpeech.SpeakAsync(mystring)} (using Xamarin.Essentials) \\
    \cline{2-4}
    & $\#15$ &
    \code{Speech.isSpeakingAsync} (using Expo SDK) &
    \rc \\
    \cline{2-4}
    & $\#17$ &
    \begin{tabular}[c]{@{}l@{}}
        \code{setDefaultRate}, \code{setDefaultPitch} (using react-native-tts)\\
        \code{Speech.speak(text, \{ pitch: 1, rate: 1, language: 'en' \})}\\ (using Expo SDK)
    \end{tabular} &
    \code{SpeechOptions} \\
    \hline
    \hline
    
    \multirow{4}{*}[+2.5ex]{\rotatebox[origin=c]{90}{A. T.}}
    & $\#18$ &
    \rc & \code{AutomationProperties.SetIsInAccessibleTree} \sm{}\\
    \hline\hline
    
    \multirow{4}{*}[-2ex]{\rotatebox[origin=c]{90}{Misc.}} &
    $\#21$ &
    \code{AccessibilityInfo.isScreenReaderEnabled} &
    \rc \\
    \cline{2-4}
    & $\#22$ &
    \code{<View} \code{accessibilityRole="header">} &
    \rc \\
    \cline{2-4}
    & $\#23$ &
    \begin{tabular}[c]{@{}l@{}}
        \code{onAccessibilityTap}\\
        \code{onMagicTap} \\
        \code{onAccessibilityEscape} \\
        \code{accessibilityActions} \\
        \code{onAccessibilityAction} (only iOS)
    \end{tabular} &
    \rc \\
    \hline

    \end{tabular}
\end{table*}

\end{document}